\documentstyle[aps,pre]{revtex}
\begin{document}
\draft
%
%
\title{Theoretical approach to biological aging} 
\author{R.M.C. de 
 Almeida$^1$, S. Moss de Oliveira$^2$ and T.J.P.Penna$^2$}
\address{1) Instituto de F\'{\i}sica, Universidade Federal do Rio 
 Grande do Sul\\ Caixa Postal 15051 - 91501-970 Porto
  Alegre, RS, Brazil\\
  2) Instituto de F\'{\i}sica, Universidade Federal Fluminense \\
Av.   Litor\^anea, s/n$^o$ - 24210-340 Niter\'oi, RJ, Brazil}
\date{\today}
\maketitle
\begin{abstract}
  We present a model for biological aging that considers the number of
  individuals whose (inherited) genetic charge determines the maximum
  age for death: each individual may die before that age due to some
  external factor, but never after that limit. The genetic charge of
  the offspring is inherited from the parent with some mutations,
  described by a transition matrix. The model can describe different
  strategies of reproduction and it is exactly soluble. We applied our
  method to the bit-string model for aging and the results are in
  perfect agreement with numerical simulations.
\end{abstract}

%

\section{Introduction}
Aging is an extremely complex biological phenomenon of immense
importance and interest. Recent progress in studying aging (or
senescence) has pointed out the importance of both genetic and
environmental components~\cite{rose,charlesworth,pb}.  Although no
single theory fully explains all aspects of the aging phenomena, the
evolutionary and free radical theories, in particular, are supported
by significant observational and experimental evidence.  Evolutionary
explanations of aging fall into two classes.  First, according to the
optimality theory, organisms might have evolved the optimal life
history, in which survival and fertility late in life are sacrificed
for the sake of early reproduction. Second, the life history might be
depressed below this optimal compromise by the influence of
deleterious mutations; since selection against late-acting mutations
is weaker, deleterious mutations will accumulate, i.e., impose a
greater load  late in life.

Besides experiments with flies and data from human populations,
computer simulations are a widely used tool to study
aging~\cite{pb,bjp,americo,book}. The recently introduced bit-string
model~\cite{jsp}, based on the mutation accumulation
hypothesis~\cite{rose,charlesworth,pb}, is able to reproduce the
exponential increase of the mortality with age~\cite{zeits} - this
behavior is known as the Gompertz's law.  Some applications of this
model have been reviewed\cite{americo,bjp2} for both asexual and
sexual versions, with emphasis on computer simulations.  However, only
a few analytical results on this model are available.  Namely the
exact results for the survival rates\cite{salmon1,salmon2} and
dynamical aspects~\cite{roux} of the catastrophic senescence of the
Pacific salmon and the description of stationary states through Leslie
matrices~\cite{tole}. In this work, we present analytical results for
a general asexual model for different strategies of reproduction. Even
for the special cases where analytical results for the bit string
model are available, our techniques are more easily and efficiently
implemented. This paper is organized as follows: in the next section,
we present a general formalism to biological aging.  In section 3, we
review the bit-string model and build the mutation matrix for it. In
section 4 we present comparisons with computer simulations. Finally,
we present our conclusions.

\section{The model}

Consider a population with $N(t)$ individuals living in an
environment with finite resources at time $t$, and with diverse
genetic charges which determine different limit ages $m$, beyond
which the individuals cannot survive.  Suppose the natural
resources allow a maximum population $N_{max}$, to be considered in
a logistic (Verhulst) factor. We note $x(a,m,t)$ as the relative number of
individuals with age $a$ and programmed death age $m$ at time $t$,
that is,
\begin{equation}
\label{xam}
x(a,m,t)=\frac{N(a,m,t)}{N_{max}}   \;\;  .
\end{equation}
Also, consider the initial and final reproduction ages as $R$ and
$R_f$. The evolution of the population in a discrete time is
described by
\begin{eqnarray}
\label{evo}
x(a+1,m,t+1) &=& \left[ 1-x(t) \right] x(a,m,t) \;\;\;\;\;
\mbox{for } 1 \leq a < m-1 \nonumber \\
x(a+1,m,t+1) &=& 0  \;\;\;\;\;\; \mbox{for } a \geq (m-1) \\
x(1,m,t+1) &=& b[1-x(t)] \sum_{m'} A_{mm'} \sum_{a=R}^{R_f}
x(a,m',t) \;\;\; , \nonumber 
\end{eqnarray}
where
\begin{equation}
\label{xt}
x(t)=\sum_{m} \sum_{a=1}^{m-1} x(a,m,t) \;\;\; ,
\end{equation}
$b$ is the number of offspring in each reproduction and $A_{mm'}$
is the birth matrix, that gives the probability of an $m'$-parent
having an $m$-offspring. When no mutation is allowed, $A_{m m'} =
\delta_{m\;m'}$. We also considered $a=1$ to be the first year of
life, so that $a\geq 1$. Note that the maximum age $a$ reached by an
individual is $m-1$.

Eqs.(\ref{evo}) describe the evolution of populations whose
transmitted genetic charge contains the information about the
maximum age of death; the way this genetic charge may change from
parent to offspring through  mutations as well as how the maximum age of
death is distributed among the population depends on further
details, typical of each species or theoretical model.  The
solution to equations (\ref{evo}) strongly depends on the form of
the birth matrix $A$. 


Hence, to completely investigate the evolution problem given by
equations (\ref{evo}), we must first choose a transition matrix A.
For that we consider in the next sections the bit-string model for
biological aging, that seems to have grasped some of the
interesting features of the age structure of populations.

On the other hand, when this transition matrix satisfies some general
features, the present model is exactly soluble, and before applying it to the
bit-string model, we present this exact {\it stationary} solution.

From eqs.(\ref{evo}) one obtains
\begin{equation}
\label{nam}
x(a,m)=\left( 1-x\right)^{a-1} x(1,m) \hspace{1cm} \mbox{for  }
1<a< m   \;\;\; , 
\end{equation}
where we explicitly assume the stationary solution and from now on
will not write the time dependence. Our methods thus do not deal with
mutational meltdown where the population is doomed to extinction.

We consider the stationary solution and the mutation matrix $A_{mm'}$ to
fulfill the following assumptions:
\begin{enumerate}
\item $x(1, m\rightarrow \infty)=0$, what is observed in living
  populations since offspring with unlimited expected life length are
  not possible (in the bit-string model this condition is equivalent
  to $x(1,B+T)=0$, where $B$ is the bit-string length and $T$ the
  lethal number of accumulated diseases);
  
\item $A_{m m'}$ is a triangular matrix such that $A_{m m'}=0 \; $ for
  $m>m'$, that is, parents cannot give birth to offspring with larger
  life expectancy (which corresponds to only bad mutation in the
  bit-string model). This is not biologically unrealistic for well
  adapted populations - advantageous mutations are expected to be
  extremely rare due to the large times required for noticeable
  species evolution;
  
\item $A_{m m} \neq 0$, that is the probability that the parent gives
  birth to offspring with the same expected life length is different
  from zero. This condition is also expected in biological populations
  and
\item $A_{mm} < A_{m'\;m'}$ if $m>m'$, the probability that a parent
  gives birth to offspring with the same expected life length $m$
  decreases with $m$. In other words, the larger the parent expected
  life length is, the larger the probability that a difference in the
  genetic charge of the offspring effectively reduces their expected
  life length ( the better the genetic code, the larger the number of
  events that can spoil it).
\end{enumerate}

The solutions to eqs.(\ref{evo}) depend on the reproduction features
of the populations. We consider two cases: the first, when the
individuals reproduce only once, that is $R_f=R$, and the second case,
when $R_f>R$ and the individuals reproduce every year after the
reproduction age, until they die.

\subsection{Case $R_f$=$R$} 

For semelparous populations (that reproduce only at age $R$), the
relative number $x(1,m)$ of $m$-babies can be written as
\begin{equation}
\label{n0m}
x(1,m)=b (1-x)\sum_{m'} A_{mm'} x(R,m') \;\; .
\end{equation}

We first observe that as $x(a,m)=0$ for $a \geq m$ the individuals with
programmed death age $m$ below $R+1$ do not reproduce.
Nevertheless these individuals keep being born from parents with $m
\geq R+1$, due to mutations.

In a population we assume a maximum life expectancy $m=\nu$ (in the
bit-string model discussed in the next section, $\nu$ can be
taken as $B+T$) and we assume $x(1,m)=0$ for $m>\nu$. Now we can write
eq.(\ref{n0m}) for $m=\nu$:
\begin{equation}
x(1,\nu)=b \left( 1-x\right)^{R}
A_{\nu \nu} x(1,\nu) \;\; ,
\end{equation}
 since we are assuming a triangular matrix and $x(1,m)=0$ for $m>\nu$.
 Hence, either $x(1,\nu)=0$ or $b\left( 1-x\right)^{R} A_{\nu \nu}
 =1$. Now, the expression for $m=\nu-1$ reads
\begin{equation}
\label{n0bt2}
x(1,\nu-1) = b\left(1-x\right)^{R}  \left[ A_{\nu-1\,\nu-1}
x(1,\nu-1)  + A_{\nu-1\,\nu} x(1,\nu)\right] \;\; .
\end{equation}
If $x(1,\nu)\neq 0$, then $b( 1-x)^{R}=A_{\nu\nu}^{-1}$ and the
above equation reduces to
\begin{equation}
\label{x1m1}
\left( 1- \frac{A_{\nu-1\,\nu-1}}{A_{\nu\nu}} \right)
x(1,\nu-1)=\frac{A_{\nu-1\,\nu}}{A_{\nu\nu}} x(1,\nu) \;\; , 
\end{equation}
which is not possible because $x(1,m) \geq 0$ for all $m$ and the left
hand side is negative as a consequence of $A_{mm} < A_{m-1\, m-1}$
(assumption 4).  Hence $x(1,\nu)=0$ and the solutions to
eq.(\ref{n0bt2}) are either $x(1,\nu-1)=0$ or $b ( 1-x)^{R}
A_{\nu-1\,\nu-1} = 1$. This situation repeats on and on up to $m=R+1$,
such that $x(1,m)=0$ for $m >R+1$.

The equation for $m=R+1$ leaves us with two possibilities: either
$x(1,R+1)=0$ or
\begin{equation}
\label{b}
b( 1-x)^{R} A_{R+1\;R+1} =1 \;\; .
\end{equation}
The first solution reduces to no population at all, since for $m<R+1$
individuals cannot reproduce. On the other hand, the expression for
$x(1,m)$ when $m<R+1$ no longer presents the diagonal term and hence
does not give place to non-positive solutions, that is,
\begin{equation}
\label{n0m2}
x(1,m)= \frac{A_{m\,R+1}}{A_{R+1\,R+1}} x(1,R+1)\;\; , \hspace{1cm}
\mbox{for  }  m <R+1 \;.
\end{equation}
Using eqs.(\ref{xt}) and (\ref{b}), we can obtain $x(1,R+1)$:
\begin{equation}
\label{n0r}
x(1,R+1)= \frac{x^2 A_{R+1\,R+1}}{\sum_{m}^{R+1} A_{m\,R+1}
\left[1-(1-x)^{m-1} \right] } \;\;.
\end{equation}
 From eq.(\ref{b}) the total population may be easily obtained:
\begin{equation}
\label{xf}
x=1-\left(\frac{1}{bA_{R+1\,R+1}}\right)^{1/R} \;\; ,
\end{equation}
and a critical number of offspring per reproduction $b_c$ may be
defined from eq.(\ref{xf}):
\begin{equation}
\label{bc}
 b_c=\frac{1}{A_{R+1\,R+1}}   \;\; .
\end{equation}
If $b<b_c$ the only solution is $x=0$.  The population age structure 
is given by the relative number $x(a)$ of individuals
at age $a$:
\begin{equation}
\label{na}
x(a)=(1-x)^{a-1}  \sum_{m=a} x(1,m) \;\; . 
\end{equation}
The catastrophic senescence effect~\cite{salmon1,salmon2} is clearly present,
since eq.(\ref{na}) implies that $x(a)=0$ for $a>R$ since $x(1,m)=0$
for $m>R+1$.
Hence the observable quantities $x$ and $x(a)$ of the stationary
solution are explicitly calculated by the transition matrix $A$ and the
parameters $b$ and $R$.

The birth matrix $A_{mm'}$ is a central point of the model. As
stated before, it describes the probability of an $m'$-parent to
give birth to an $m$-offspring, and hence it is closely related to
the probability of mutations during the birth process. However,
assuming that the birth matrix is triangular and fulfills the basic
assumptions $2$, $3$ and $4$ for a single mutation, then it is
triangular and fulfills the same assumptions also for $M$ multiple
mutations, since in this later case the resulting birth matrix is
the product of $M$ single mutation birth matrices.  In other words,
if the above results apply to a model with a single mutation per
birth, they also apply to $M$ multiple mutations, provided that the
birth matrix is taken as the product of $M$ single mutation
matrices.

A special case happens when no mutation at all is considered. In
this case $A_{mm'}=\delta_{mm'}$ does not fulfill the basic
assumptions but the problem is still exactly soluble. In this case,
\begin{equation}
x_{M=0}=1-\left( \frac{1}{b}\right)^{1/R}
\end{equation}
and the age structure of the population is determined by the values assumed
for  $x(1,m)$, that must only satisfy
\begin{equation}
x=\sum_{m}^{R+1} \frac{1-(1-x)^{m-1}}{x} x(1,m)\;\; .
\end{equation}

\subsection{ Case $R_f>R$}

It is interesting to investigate what happens when the individuals may
reproduce more than once, that is, for $a \geq R$. We will first
consider that an individual reproduces from age $R$ until its death
(iteroparous population). In this case, birth is described by the
equation
\begin{equation}
x(1,m) = b(1-x) \sum_{m'} A_{mm'} \sum_{a=R}^{m'-1} (1-
x)^{a-1} x(1,m')  \;\;\;.
\end{equation}
Summing over $a$ we have
\begin{equation}
\label{x1m2}
x(1,m)  = b (1-x)^{R} \sum_{m'=R+1} A^*_{mm'} x(1,m') \;\; 
\end{equation}
where
\begin{equation}
A^*_{mm'}= A_{m m'}\frac{ 1-(1-x)^{m'-R}}{x} \;\;.
\end{equation}
It is straightforward to verify that since $x <1$ the renormalized
birth matrix $A^*$ does not necessarily conserve the required
properties for a catastrophic senescence solution, that is, $A^*_{m   m}$
could increase with $m$. However in order to have $x\neq 0$
there must exist some $m_{\ell}$ such that $x(1,m_{\ell}) \neq 0$
and $x(a,m)=0$ for $m>m_{\ell}$.  If $m_{\ell} > R+1$, that is,
the senescence is not catastrophic, then the following must hold:
\begin{equation}
\label{ratio}
\frac{ A_{m_{\ell}-1 \, m_{\ell}-1}}{A_{m_{\ell}\,m_{\ell}}}\;
\frac{[ 1-(1-x)^{m_{\ell}-1-R}]}{ [1-(1-x)^{m_{\ell}-R}]} <1\;\;,
\end{equation}
so that the equation for $x(1, m_{\ell}-1)$ has a non null solution,
in analogy to eq.(\ref{x1m1}). The ratio between the matrix terms is
larger than one, and so the second ratio in the left hand side of the
above equation must be small enough to compensate.  Given an
$m_{\ell}>R+1$ that satisfies eq.(\ref{ratio}), the total population
is obtained from the solution of
\begin{equation}
\label{xts}
 \frac{ (1-x)^R-(1-
x)^{m_{\ell}}}{x}=\frac{1}{b A_{m_{\ell}\,m_{\ell}}} \;\; . 
\end{equation}
Observe that when $m_{\ell}=R+1$, the above equation reduces to the
result obtained in the previous section with $R_f=R$, as it should.
  
It is  possible to obtain the maximum $m_{\ell}$ for a
given birth matrix $A$, that is,  the
maximum death age of an iteroparous population. It is useful
then to define a maximum
death age limit $\mu$:
\begin{equation}
\mu={\rm max}\{m_\ell\}\;\;\; . 
\end{equation}
where all $m_{\ell}$ satisfy eqs.(\ref{ratio}) and (\ref{xts}).  When
starting from an initial population containing death ages larger than
$\mu$, the solution converges to $m_{\ell}=\mu$.  However, depending
on the structure of the birth matrix, it can happen that
eq.(\ref{ratio}) never holds for $m_{\ell}>R+1$. In this case, the
system organizes itself as a semelparous population and all results of
the previous section apply. Also, when the population initially
contains only individuals with death age $m<\mu$, then $m_{\ell}<\mu$, 
because the birth matrix $A$ is triangular.

An intermediary case, when $R+1<R_f< \mu$, may also be considered. 
In this case the solution is $m_{\ell}=R_f+1$ for initial
conditions where $m>R_f+1$ are present.

The solution to the problem is complete with the values of $x(1,m)$
for $R+1\leq m \leq m_{\ell}$, that are obtained by finding the
solution to the eigenvalue equation
\begin{equation} 
\label{auto} 
x(1,m)=\frac{1}{A_{m_{\ell}\,m_{\ell}} [1-(1-x)^{m_{\ell}-R}]}
\sum_{m'=R+1}^{m_{\ell}} A_{m\,m'}[1-(1-x)^{m'-R}]  x(1,m') \;\;\;
, \end{equation} 
for $R+1\leq m\leq m_{\ell}$,
that satisfies the normalization condition given by
\begin{equation}
x=\sum_{m} \sum_{a=1}^{m-1} x(a,m)
\end{equation}
where $x(a,m)=(1-x)^{a-1} x(1,m)$.

Here we can also discuss the case where mutations are absent. As
before, $A_{mm'}=\delta_{mm'}$. The eigenvalue equation reduces to
\begin{equation}
x(1,m)=b \frac{(1-x)^R-(1-x)^m}{x} x(1,m) \;\;\; \mbox{for }m>R.
\end{equation} 
That means that for every $m^*>R$ there is a different solution, where
the population consists only of $m^*$-individuals and the total
population $x$ satisfies the above equation for $m=m^*$. The final state
depends on the initial conditions and the system always converges to
a solution with $m_{\ell}$ given by the largest $m$ present in the
initial population.

In the next sections we apply our approach to the bit-string model.

\section{The bit-string model}

The bit-string model for biological aging \cite{jsp} consists in a
population of $N(t)$ individuals at time $t$, each one represented
by a bit-string of $B$ bits and subject to aging, reproduction and
death. The bit $S_i$ $(i=1, \ldots, B)$ of the string contains the
genetic information of the programmed health status of the
associated individuals at age $i$: when $S_i=1$ a {\it genetically
programmed} disease is acting for all $a\geq i$.  Each individual
in the population may survive up to $T$ of these diseases, that is,
it cannot survive longer than age $m$, at which the sum of bits
from zero to $m$ is $T$. Individuals cannot reproduce before age
$R$, when it gives birth to $b$ offspring either once
(semelparous) or each year (iteroparous) until death. An offspring
has the same string as the parent except for $M$ bits that are
randomly changed, to simulate genetic inheritance and mutation. The
environment limitations are taken into account through a Verhulst
factor, that reduces the number of individuals by $(1-N/N_{max})$,
where $N_{max}$ is the maximum allowed population size. From now on we
shall discuss in terms of the relative population quantities
$x=N/N_{max}$, $x(a)=N(a)/N_{max}$ and $x(a,m)=N(a,m)/N_{max}$. 

Exact results have been available only for semelparous populations with
$T=1$ and deleterious mutations\cite{salmon1}. The analytical approach
to aging that we present in eqs.(\ref{evo}) may be adapted to
reproduce the simulation model.  To account for the finite length of
computer bit-strings, we also take $x(a,m,t)=0$ for
$a>B$ whereas $m \geq T$, since in the bit-string model no individual
may have its programmed death age below $T$. Moreover, $m \leq
(B+T)$ where $m=B+k$ is associated to individuals with only $k<T$ bits
equal to 1.
 
To describe the bit-string model for biological aging, we note that
the probability of finding an individual with programmed death age $m$
is equivalent to the probability of finding the $T^{th}$ genetically
programmed disease acting on age $m$, or to find the $T^{th}$ bit equal to
unity in the $m^{th}$ position of the bit-string associated
to the individual. Also, the birth matrix $A$ should reflect the
change in the death age implied by the flipping of randomly chosen
bits when building an offspring bit-string.  In a recent work, Ito
\cite{ito} obtained an analytical approach to the bit-string model by
considering the relative number $n(\vec{S},a,t)$ of different
bit-strings $\vec{S}=(S_1,S_2,\ldots,S_B)$ present in the population.
The present model reproduces the evolution equations of Ito~\cite{ito}
by summing over all bit-strings with the same death age $m$, that is
$x(a,m,t)=\sum_{\{\vec{S}\} }n(a,\vec{S},t) \delta(\ell-m)$ where
${\ell}$ is the position (locus) of the $T^{th}$ inherited disease in the
bit-string $\vec{S}$. Due to the form of the birth
matrices of the bit-string model, the general calculations performed
in the previous sections still apply and exact results may be
produced.

\subsection{The mutation matrix}

To write the birth matrix $A_{m\;m'}$ for the bit-string model, we
first consider the one-mutation matrix $F$, when at most one bad
mutation happens at birth in the bit-string model: a random site is
chosen and set to one, regardless its previous state. The mutation
matrix is triangular and reads
\begin{eqnarray}
\label{mb}
\mbox{for  } m'\leq B: & & \nonumber \\
F_{mm'} &=& \frac{T}{B} \frac{(m'-T)!\;(m-1)!}{(m-T)!\;(m'-1)!}
\hspace{1.5cm}
\mbox{for } m\leq m'-1 \nonumber \\
F_{mm'} &=& \frac{B-m'}{B}+\frac{T}{B} \hspace{1.5cm} \mbox{for  }
m=m' \nonumber \\ 
F_{mm'}&=& 0 \hspace{2.0cm}\mbox{for  } m'+1\leq m \leq B+T
\nonumber \\
\mbox{ for  } m'= B+1 : & & \nonumber \\
F_{mm'} &=& \frac{T}{B} \frac{(m-1)!\;(B-T+1)!}{(m-T)!\; B!}
\hspace{1.5cm} \mbox{ for  } T\leq m \leq B \nonumber \\
F_{mm'} &=& \frac{T-1}{B} \hspace{2.0cm} \mbox{for  } m=B+1
\nonumber \\ 
F_{mm'} &=& 0 \hspace{2.2cm} \mbox{for   } m>B+1 \nonumber \\
\mbox{for  } B+2 \leq m' \leq B+T \nonumber \\
F_{mm'} &=& 0 \hspace{2.2cm} \mbox{for   } m<m'-2 \nonumber \\
F_{mm'} &=& \frac{m'-T}{B} \hspace{2.0cm} \mbox{for } m = m'-1
\nonumber \\
F_{mm'} &=& \frac{B+T-m'}{B} \hspace{2.0cm} \mbox{for }
m=m'\nonumber \\ 
F_{mm'} &=& 0 \hspace{2.2cm} \mbox{for } m'+1 \leq m \leq B+T
\nonumber  
\end{eqnarray}
The matrix elements are obtained considering that the probabilities of
finding a bit one in different ages before age $m$ are not correlated.
Now, the string length $B$ and the maximum number of diseases $T$ of
the bit-string model have been already taken into account (actually
$B$ is necessary only due to the finite limits of computers.

In the case that more than one mutation may happen at birth, say $M$
mutations, the mutation matrix is taken to the power $M$, that is
$A=F^M$. The birth matrix is triangular and considering $R_f=R$
populations, it fulfills the required conditions for a catastrophic
senescence for any $M>0$.

\section{Results}

Considering the bit-string model, there are three different ways of
obtaining results: analytically as presented in the previous
sections, numerically iterated solutions to eqs.(\ref{evo}) and
numerical simulations. In what follows we will discuss these three
forms, for each case we have considered.

We first present the results for $R_f=R=11$ (age at reproduction of
the Pacific salmon, in years). The birth matrix $A$ is obtained from
the one mutation matrix $F$, that is
\begin{equation}
  A=F^M \;\;\; \mbox{for } M>0 \;\; ,
\end{equation}
and $A_{mm'}=\delta_{mm'}$ for $M=0$.  In fig. 1 we present a snapshot
of the time evolution of $x$ for $M=1$ and $T=1$, considering
different initial conditions: a) fixed point solution as given by
eq.(\ref{xf}) and b) all individuals with age 1 and $m=32$, i.e, free
of mutations. The iteration of eq.(\ref{evo}), starting from the
solution of eq.(\ref{xf}) confirms the existence of a fixed point of
the dynamics where the system remains for all times. However, for any
values of $M$ and $T$, if $b>b_c$ and $R_f=R$ we also found
oscillatory states with period $11$ ($=R$).  These oscillations may
vary in amplitude and phase, depending on the initial condition. They
have been previously observed \cite{roux,thom} and are due to the
non-overlap of generations in semelparous populations \cite{roux}.
The effect of increasing $T$, the limit number of diseases that kill
the individual, is to increase the population. On the other hand,
increasing $M$ decreases the total population, because offspring with
low $m$ are more frequently generated in each birth process.
Simulations show that the fixed point solution is very unstable: any
perturbation drives the system slightly away from the fixed point,
after a few steps the population is driven to an oscillatory regime.
Another point worth noting is the extremely slow convergence of the
age distribution. When the population size reaches the equilibrium
(after less than 2000 steps) the population with age one year after the age
at reproduction has not vanished. An exponential fitting ($x(a)= A
\exp(\alpha t)$) to this age gives a small valued exponent
$\alpha=-0.004$ which guarantees that the population older than age at
reproduction will eventually vanish.

In fig. 2 we present the survival rates, defined as
\begin{equation}
S(a) = \left( \frac{x(a+1)}{x(a)} \right) \;,\;\; a\geq 1\;\;\;\; . 
\end{equation} The catastrophic senescence effect is clearly
present in this figure.  The excellent agreement between computer
simulations and the analytical results can be also seen.

We consider now $R<R_f$, for the bit-string model.  Considering $B=64$
and reproduction every year until death, our results do not show any
oscillatory behavior: regardless the initial conditions the
system converges to a fixed point solution.  In fig.  3 we present a
semi-log plot of the mortality rate at age $a$ defined as\cite{azbel}
\begin{equation} 
q_a = - \ln \left( 1- \frac{S(a)}{S(1)}\right) ,
\end{equation} 
for $M=1$,$T=1$,$b=0.1$. This normalization has been proved to
eliminate the Verhulst factor influence\cite{salmon2}.  Using this
normalized mortality curve we can see the Gompertz's region (from ages
10 to 25). A clear change in the behavior at the minimum age of
reproduction $R$ - where we expect that mutation accumulation effects
are not relevant - can be seen in this figure.  Another deviation
occurs also at older ages. The existence of several Gompertz's regions
was proposed by Gompertz~\cite{azbel} and it was already studied in
the bit-string framework~\cite{racco}. To obtain the analytical
results for the asymptotic solution we must first obtain the maximum
death age of the population. For these parameters, the maximum death
age is $\mu = 33$. Starting from an initial condition where $m>33$ are
present, the population stabilizes with $m_{\ell}=\mu=33$ (maximum
age=$m_{\ell}-1$).  Solving eq.(\ref{xts}) for the adequate birth
matrices $A$, we can find the asymptotic total population. Again we
present computer simulations for comparison.  After an enough number
of time steps, individuals with advanced ages ($a>33$) tend to
disappear from the population.  Our results were obtained from longer
series ($3\times 10^6 time steps$) than than the ones from 
intensively parallel simulations by Meisgen~\cite{meisgen} (800,000
time steps).

It can be argued about the sizes of the simulated systems.  This point
is one of the important advantages of the present treatment. We
adopted, in computer simulations, population sizes around 300000
individuals (considering $B=32$ for comparisons, which is the most
used value in computer simulations). In a Pentium 150MHz (32Mb RAM)
running Linux and using a very optimized code, 1025 sec are needed to
simulate 20000 time steps. However, solving the equations for this
same time interval, using the present approach only just 20 sec are
needed, that is, in our strategy is three orders of magnitude faster
than a simulation (for 300000 individuals)! Also, it is worth
remarking that no population finite size effects are present in this
treatment. (The finite size effects related to the bit string length
are exactly taken into account).  Moreover the storage needed to
simulate it is basically $2*B \times B \times 8$ bytes (considering
double precision float point numbers) whereas a $N=300000/8$ bytes
were required to the computer simulations (using bitwise operations).
However, we have to emphasize that in the current approach the
computer time increases as $B^2$ whereas in computer simulations the
CPU time increases linearly with $B$. 

In summary, the analytical calculations agree perfectly well with both
the stationary solutions found by numerically iterating
eqs.(\ref{evo}) and with simulations. For $R=R_f$,
oscillatory stable solutions may also be found when solving
iteratively the evolution equations.  Finally, we observe that the
computer time required for numerically iterated solutions for the
analytical evolution equations is very short due to the size of the
transition matrices ($B \times B$), in comparison to previous models
where the birth matrices are $2^B \times 2^B$ \cite{ito}.  This CPU
time required in the present model is still orders of magnitude
smaller than that required by computer simulations of systems
comparable to real sizes, with the traditional $B=32$ age intervals.

\section{Conclusions}

We have presented a theoretical model to describe how inherited
genotypes may determine the age structure of a population. This model
takes into account the inherited maximum lifespan, which is passed
from parent to offspring according to a birth matrix that may
contemplate the possibility of mutations. When this birth matrix is
triangular, the stationary solution is obtained analytically for
semel- and iteroparous population. Considering that reproduction
occurs for $R\leq a \leq R_f$, the expected catastrophic senescence is
obtained for $R_f=R$, but it can also occur for $R_f\rightarrow
\infty$, depending on the structure of the birth matrix.  In general
cases, successive iterations (not simulations) always reach asymptotic
solutions to the evolution equations that does not require large
computational resources.  Applying this technique to the bit-string
model, we found that iteroparous population always stabilize at the
theoretically predicted stationary solution, but semelparous ones may
also present stable oscillatory behavior with period $R$ which
amplitudes depend on the initial conditions. The agreement with
simulation results are noticeable. This approach is at least three
orders of magnitude faster than computer simulations and still more
memory saving. It is worth to remember that computer simulations have
been used up to now, as the most important tool to aging studies.
Therefore, we believe that our results can be useful for further
studies on biological aging.  A natural extension is to consider a
continuous time limit, where the time scale is given by $R$, the
biological relevant quantity.  This extension is now in progress and
will be presented in due time.

\section{Acknowledgement}

This work has been partially supported by Brazilian agencies CNPq,
CAPES, FINEP, FAPERGS and FAPERJ. We acknowledge valuable suggestions and
fruitful discussions with Dietrich Stauffer, Cristian Moukarzel,
P.M.C. de Oliveira,  J.C.M. Mombach and J.S. S\'a Martins.

\newpage

\section{Figure Captions}

Figure 1. A snapshot of the time evolution of the population for two
different initial conditions: dotted line refer to the fixed point
solution of eq.(\ref{xf}) and solid lines to all individuals initially
free of mutations. The parameters used in both conditions were:
$R_f=R=11$, $b=10$, $M=T=1$.  From the top to bottom the solid lines
mean the total population, population at ages 1,5 and 12 years. We can
note a displacement of the peaks at each age curve. Since the total
population is the sum of all ages, these peaks present a smoother aspect. 
Population vanishes for  ages above $R=11$. However, the decay
of these curves for the oscillatory solutions is extremely slow.

\vskip 2\baselineskip

Figure 2. Survival rates $S(a)$, in units of $S(1)$, for $R_f=R=11$,
$b=10$, $M=T=1$.  Catastrophic senescence is clearly seen in this
plot. Solid lines are the analytical results and the full circles are
the results from computer simulations.

\vskip 2\baselineskip

Figure 3. Normalized mortality rate $q(a)$ for $R=8$, $R_f=64$,
$M=T=1$, and $b=0.1$. The full line refers to the theoretical
predictions after 3 million time steps (this is the longest time series
obtained for the model). Computer simulation results after $10^4$
steps are represented by circles. The population beyond age $33$ will
eventually disappear (after one million years, according to the
theoretical predictions).

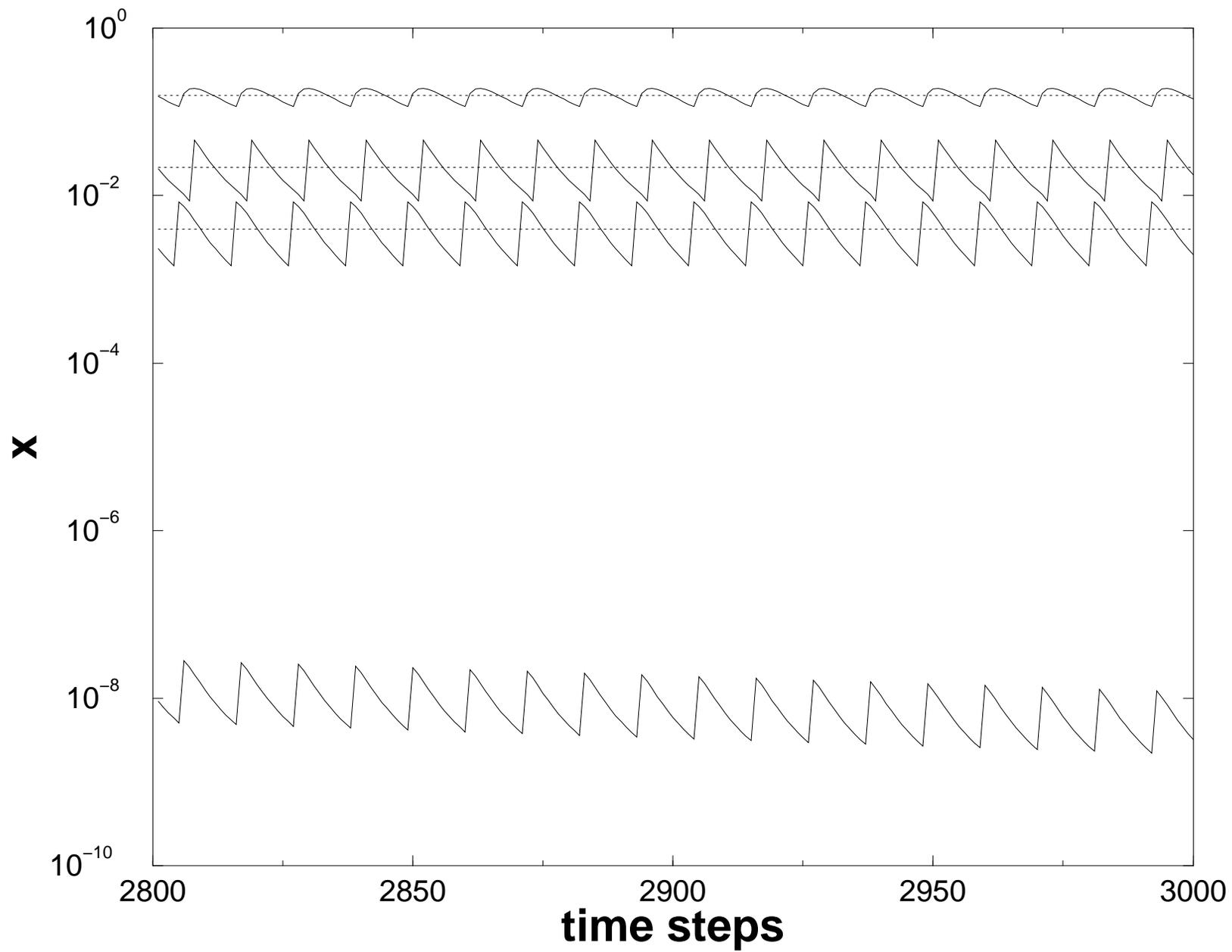
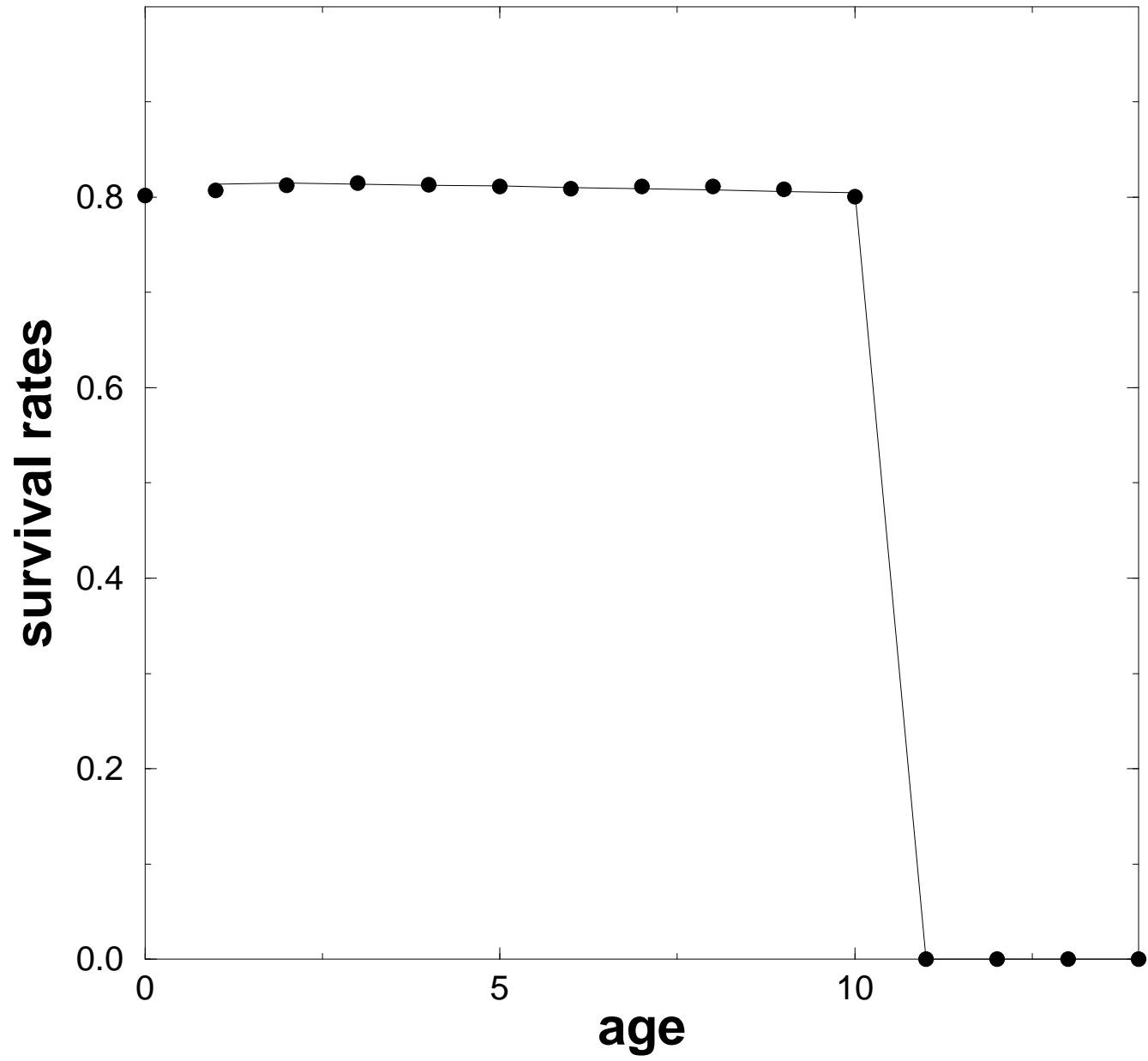
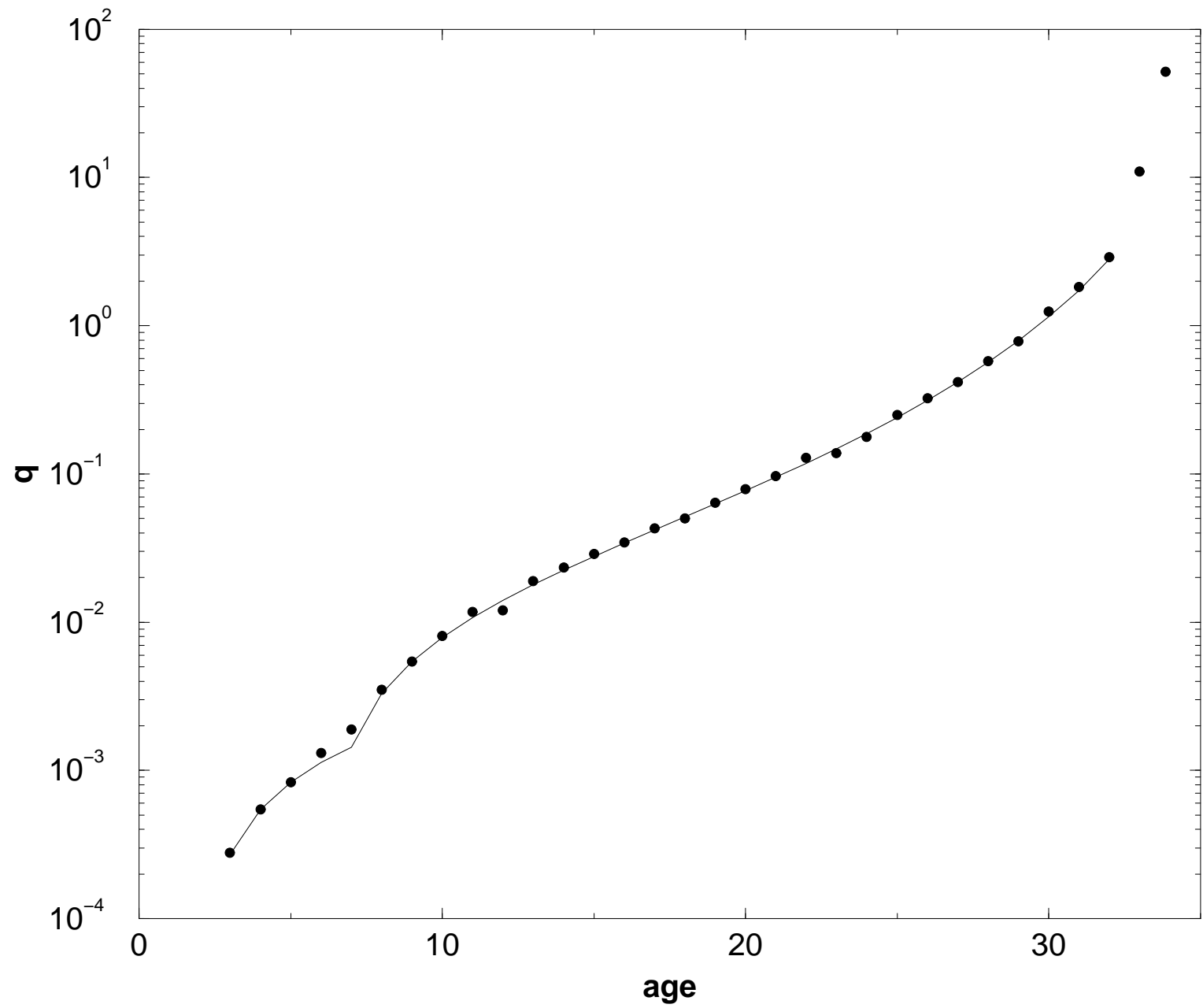
\end{document}